\documentclass[sn-mathphys-num,ii]{sn-jnl}

\usepackage{graphicx}%
\usepackage{multirow}%
\usepackage{amsmath,amssymb,amsfonts}%
\usepackage{amsthm}%
\usepackage{mathrsfs}%
\usepackage[title]{appendix}%
\usepackage{xcolor}%
\usepackage{textcomp}%
\usepackage{manyfoot}%
\usepackage{booktabs}%
\usepackage{algorithm}%
\usepackage{algorithmicx}%
\usepackage{algpseudocode}%
\usepackage{listings}%
\usepackage{ulem}
\usepackage{lmodern}
\newcommand\redsout{\bgroup\markoverwith{\textcolor{blue}{\rule[0.5ex]{2pt}{1.4pt}}}\ULon}

\renewcommand{\emph}[1]{\textit{#1}}

\begin{document}

\title[""]{Measurement of the $\mathrm{{}^{12}C}(e,e')$ cross sections at $Q^2=0.8\,\mathrm{GeV}^2/c^2$}

\author[1,2]{\fnm{M.} \sur{Mihovilovi\v{c}}} 
\author[3,6]{\fnm{L.} \sur{Doria}}
\author[4]{\fnm{P.} \sur{Achenbach}}
\author[5]{\fnm{A.~M.} \sur{Ankowski}}
\author[3,6]{\fnm{S.} \sur{Bacca}}
\author[7]{\fnm{D.} \sur{Bosnar}}
\author[3,6]{\fnm{A.} \sur{Denig}}
\author[3]{\fnm{M.~O.} \sur{Distler}}
\author[3]{\fnm{A.} \sur{Esser}}
\author[7]{\fnm{I.} \sur{Fri\v{s}\v{c}i\'{c}}}
\author[8]{\fnm{C.} \sur{Giusti}}
\author[3]{\fnm{M.} \sur{Hoek}}
\author[3]{\fnm{S.} \sur{Kegel}}
\author[3]{\fnm{M.} \sur{Littich}}
\author[9]{\fnm{G.~D.} \sur{Megias}}
\author[3,6]{\fnm{H.} \sur{Merkel}}
\author[3]{\fnm{U.} \sur{M\"{u}ller}}
\author[3]{\fnm{J.} \sur{Pochodzalla}}
\author[3,6]{\fnm{B.~S.} \sur{Schlimme}}
\author[3]{\fnm{M.} \sur{Schoth}}
\author[3,6]{\fnm{C.} \sur{Sfienti}}
\author[2,1]{\fnm{S.} \sur{\v{S}irca}}
\author[3,6]{\fnm{J.~E.} \sur{Sobczyk}}
\author[3]{\fnm{Y.} \sur{St\"{o}ttinger}}
\author[3,6]{\fnm{M.} \sur{Thiel}}

\affil[1]{\orgname{Jo\v{z}ef~Stefan~Institute}, \postcode{SI-1000}, \city{Ljubljana}, \country{Slovenia}}

\affil[2]{\orgdiv{Faculty of Mathematics and Physics}, \orgname{University of Ljubljana}, \postcode{SI-1000}, \city{Ljubljana}, \country{Slovenia}}

\affil[3]{\orgdiv{Institut f\"{u}r Kernphysik}, \orgname{Johannes Gutenberg-Universit\"{a}t Mainz}, \postcode{DE-55128}, \city{Mainz}, \country{Germany}}

\affil[4]{\orgname{Thomas Jefferson National Accelerator Facility}, \city{Newport News}, \postcode{VA 23606}, \country{USA}}

\affil[5]{\orgdiv{Institute of Theoretical Physics}, \orgname{University of Wroc{\l}aw}, pl. Maxa Borna 9, \postcode{50-204} \city{Wroc{\l}aw}, \country{Poland}}

\affil[6]{\orgdiv{PRISMA$^+$ Cluster of Excellence}, \orgname{Johannes Gutenberg-Universit\"{a}t Mainz}, \postcode{DE-55128}, \city{Mainz}, \country{Germany}}

\affil[7]{\orgdiv{Department of Physics}, \orgname{University of Zagreb}, \postcode{HR-10002}, \city{Zagreb}, \country{Croatia}}

\affil[8]{\orgname{INFN, Sezione~di  Pavia},  \postcode{I-27100} \city{Pavia}, \country{Italy}}

\affil[9]{\orgdiv{Departamento~de~F\'isica~At\'omica, Molecular~y~Nuclear}, \orgname{Universidad~de~Sevilla},  \postcode{41080} \city{Sevilla}, \country{Spain}}

\abstract{
We present the findings of a study based on a new inelastic electron-scattering 
 experiment  on the ${}^{12}$C nucleus focusing on the kinematic region of $Q^2=0.8\,\mathrm{GeV}^2/{c}^2$.
The measured cross section is sensitive to the transverse response function and provides a stringent test of theoretical models, as well as of the theoretical assumptions made in Monte-Carlo event-generator codes developed for the interpretation of neutrino-nucleus experiments, such as DUNE and HyperK. We find that modern generators such as GENIE and GiBUU reproduce our new experimental data within 10$\%$.
}


\maketitle

\section{Introduction}
\label{intro}

Electrons represent a very precise probe for the investigation of the atomic nucleus~\cite{Boffi:1993gs}. In the past decades, experiments with electrons have provided increasingly accurate information on the structure of nuclei and their constituents~\cite{G0:2009wvv, LabHallA:2014wqo, PVDIS:2014cmd, Defurne:2017paw, A1:2017mxs, A1:2019qhv, Esser:2018vdp,JeffersonLabHallATritium:2020mha}. At the heart of this effort are the inelastic scattering experiments on nuclear targets at energies below $1\,\mathrm{GeV}$, which give insight into the properties and dynamics of nucleons embedded in the nuclear medium. In such scattering processes, an electron with energy $E_0$ interacts with the nucleus at rest by exchanging a virtual photon transferring energy $\omega$ and momentum $\vec{q}$, 
such that $Q^2 = \vec{q}^2 - \omega^2 >0$. Using the nucleon mass $m_N$ as a scale, the energy and momentum transfer variables can be rewritten in dimensionless form as
$$
\lambda = \frac{\omega}{2m_N}\,,\quad\vec{\kappa} = \frac{\vec{q}}{2m_N}\,,\quad \tau = \frac{Q^2}{4m_N^2} = \vec{\kappa}^2 - \lambda^2\,.
$$
The differential cross section describing the inclusive interaction of the electron with the nucleus can be written as
\begin{eqnarray}
\frac{d^2\sigma}{d\Omega d\omega} = \sigma_M \left[ v_L R_L(\omega,q) 
+ v_T R_T(\omega,q) \right]\,,           
\label{eq:crosssection}
\end{eqnarray}
where $\sigma_M$ is the Mott cross section~\cite{Boffi:1993gs}, while $v_L$ and $v_T$ are kinematic factors given by
$$
v_L = \left(\frac{\tau}{\kappa^2} \right)^2\,,\qquad
v_T = \frac{\tau}{2 \kappa^2} + \tan^2 \frac{\theta_e}{2}\,,
$$
and $\theta_e$ is the angle of the scattered electron. The cross section depends on two nuclear responses, the longitudinal and the transverse response functions, which are both functions of $\omega$ and $q=|\vec{q}|$. The longitudinal response function, $R_L(\omega, q)$, depends on the charge operator and carries information on the nucleon-nucleon correlations, while the transverse response function, $R_T(\omega, q)$, is driven by the magnetic currents~\cite{Boffi:1993gs}.

Various cross section measurements were performed, mostly before 2000, but the acquired data were dominated by the longitudinal (charge) part of nuclear response.  Historically, the most extensively studied nucleus has been carbon. For this nucleus, the richest sample of $(e,e')$ data exists. It consists of almost 3500 data points from 12 experiments~\cite{Benhar:1994hw, Ankowski:2020:PhysRevD.102.053001} for energies between $0.12\,\mathrm{GeV}$ and $17.3\,\mathrm{GeV}$ and scattering angles up to $145\,{}^\circ$.   These data have been used to study the structure of this nucleus and to develop models describing its electromagnetic response. 

In the past, theoretical calculations for ${}^{12}\mathrm{C}$ were often limited to the quasi-elastic (QE) region, and the most demanding part was the description of the transverse response, which has been for a long time incomplete~\cite{Boffi:1993gs}. A more comprehensive description of the $\mathrm{{}^{12}C}(e,e')$ cross section was developed in the microscopic calculation by Gil {\it et al.} ~\cite{Gil:1997bm}, and more recently Megias {\it et al.} ~\cite{Megias:2016lke} proposed a superscaling model, called SuSAv2-MEC, 
which considers the complete inelastic spectrum. 
The model shows quite good agreement with data over a broad range of energy transfer. However, the description of the cross sections at large scattering angles remains incomplete. 

An important motivation for new studies of inclusive cross sections comes also from the neutrino physics community \cite{e4nu, e4nu-th}. Short- and long-baseline neutrino experiments detect neutrinos through their interactions with nuclei and aim at the precise measurement of neutrino masses, mixing angles, and CP-violating phase in the lepton sector. These measurements represent one of the highest priorities of contemporary fundamental physics and hinge on the ability of the experiments to reconstruct the neutrino energy and on the precise knowledge of the neutrino-nucleus cross sections. Although a vigorous experimental program for the measurement of such cross sections is in progress~\cite{Murphy:2019wed, Dai:2018gch, Dai:2018xhi, Jiang_JeffersonLabHallA:2022cit, JeffersonLabHallA:2022ljj}, neutrino experiments are mostly limited by statistical uncertainties and the lack of knowledge of the neutrino flux. Electron scattering experiments, with a precisely determined beam energy and the possibility to perform inclusive as well as exclusive measurements with different final states, have the potential to provide very precise data for testing the nuclear models employed in neutrino experiments.

Indeed, it has been demonstrated that the interpretation of the measured neutrino oscillations requires extensive theoretical and experimental support from the nuclear physics community. In this context the ${}^{12}\mathrm{C}(e,e')$ reaction has played an important role in the development of reliable models describing cross sections~\cite{Megias:2016lke, Megias_PhysRevD.101.033003} in experiments like MiniBooNE~\cite{AguilarArevalo:2007it}, MINERvA~\cite{Stancil:2012yc}, and T2K~\cite{Abe:2011ks} that use carbon-based materials (mineral oils, plastic scintillators) as detector medium. To ensure further involvement of modern neutrino event generators like GENIE~\cite{Andreopoulos:2009rq} and GiBUU~\cite{Buss:2011mx}, the advances in the built-in theoretical models must be complemented by the new experimental data on relevant nuclear targets and in relevant kinematics~\cite{PhysRevD.103.113003, CLAS:2021neh}. 

In this paper we focus on $^{12}$C
and present new data at $Q^2=0.8\,\mathrm{GeV}^2/c^2$ for two reasons. On the one hand carbon is an interesting target for neutrino experiments as mentioned above, and on the other hand new data at large $Q^2$ in a kinematic dominated by the transverse response will boost further theoretical progress~\cite{Cloet:2015tha,Lovato12C_el,Lovato12C_nu,Bacca:2006ji,Saori_STA,Sobczyk:2020,Sobczyk_PRL,Saori_A3, Sobczyk_SF_4He, Sobczyk_RT_16O,Sobczyk_2024_SF_16O}.

\section{Experiment}
\label{sec:analysis}

The measurement of the inclusive cross section on ${}^{12}\mathrm{C}$ was performed at the Mainz Microtron (MAMI) facility using the spectrometer setup of the A1 Collaboration~\cite{Blomqvist}. In the experiment, an electron beam with energy $E_0=855\,\mathrm{MeV}$ was used in combination with a $43\,\mathrm{mg/cm^2}$ thick carbon foil target. For measuring the cross section as a function of energy of scattered electron $E'$ we employed a magnetic spectrometer (spectrometer A) with $20\%$ momentum acceptance and 28~msr angular acceptance. The spectrometer was positioned at a fixed angle of $70^\circ$, while its momentum settings were adjusted to measure the cross section as a function of $\omega = E_0-E'$. The measurements were made for seven different momentum settings between $310\,\mathrm{MeV}/c$ and $650\,\mathrm{MeV}/c$ in order to collect data in the region of the QE peak and the $\Delta$-resonance. For each setting we collected $1.8$ million events. The central momentum of each setting was measured to a relative accuracy of $8\times10^{-5}$. The spectrometer was equipped with a detector package consisting of two layers of vertical drift chambers (VDCs) for tracking, two layers of plastic scintillation detectors for triggering, and a threshold Cherenkov detector for electron identification. The beam current was between $2\,\mu\mathrm{A}$ and $3\,\mu\mathrm{A}$ and was limited by the maximum data acquisition rate, resulting in a raw rate of about $500\,\mathrm{Hz}$. The current was determined by a non-invasive fluxgate-magnetometer with an accuracy of $<0.2\,\mathrm{\%}$.

\begin{figure}[!ht]
\includegraphics[width=0.8\linewidth]{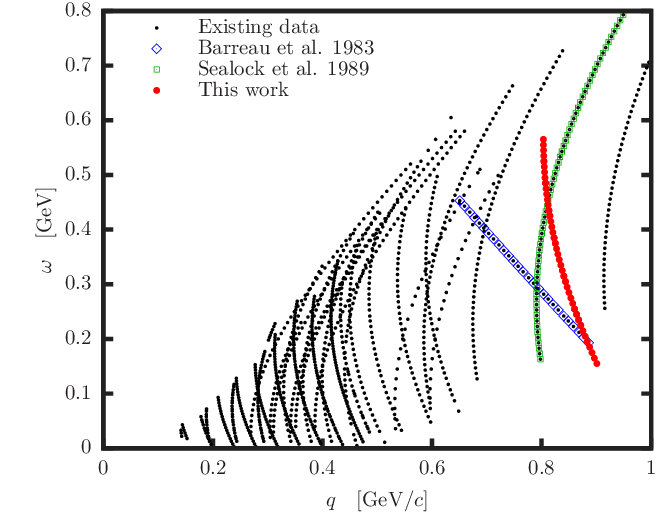}
\caption{Kinematic configurations of the ${}^{12}\mathrm{C}(e,e')$ cross section data in terms of energy and momentum transfers $\omega$ and $|\vec{q}|$. Black points represent available cross section measurements with relative uncertainties smaller than $10\,\%$. The kinematics covered by this work are presented with red circles. The complementary measurements of Barreau~{{\it et al.} }~\cite{BARREAU1983515}, and Sealock~{{\it et al.} }~\cite{Sealock1989}, also at $|\vec{q}| \sim 0.8\,\mathrm{GeV}/c$, are shown with blue diamonds and green squares, respectively. \label{fig_Kinematics} }
\end{figure}

The experiment provided new cross sections measurements in the region of beam energies and scattering angles, where the existing measurements are very sparse, see Fig.~\ref{fig_Kinematics}. The quasi-elastic peak is centered at $|\vec{q}| = 0.84\,\mathrm{GeV}/c$, thus nicely complementing previous measurements at $|\vec{q}| \approx 0.8\,\mathrm{GeV}/c$ performed at $560\,\mathrm{MeV}$ and $1299\,\mathrm{MeV}$~\cite{BARREAU1983515,Sealock1989}.

The experimental cross sections for the ${}^{12}\mathrm{C}(e,e')$ reaction were extracted from the data by dividing the measured distributions of counts by the integrated luminosity and the solid angle accepted by the spectrometer. 

The accepted solid angle was simulated using a dedicated simulation for the three spectrometers facility of the A1 Collaboration~\cite{Blomqvist}, \texttt{Simul++}. To ensure a reliable comparison with the data, the simulation included realistic momentum and spatial resolutions of the spectrometer. The relative momentum, angular, and vertex resolutions (FWHM) were $2.4\times 10^{-4}$, $4.7\,\mathrm{mrad}$, and $9.4\,\mathrm{mm}$, respectively. The simulation also considered the electron energy corrections due to multiple scattering and radiation losses. The internal and external radiative corrections were included using the formalism of Mo and Tsai~\cite{MoTsai}. The accompanying multiple scattering corrections in the target and surrounding material were approximated by a Landau distribution~\cite{landau1944}. Altogether, the energy corrections have less than $4\,\mathrm{\%}$ effect on the measured cross section.  

The integrated luminosity was determined from the product of the accumulated charge and the surface density of the target material, corrected for dead-time and DAQ prescale factors. The luminosity was determined separately for each collected data-sample to ensure that dead-time and prescale corrections were consistently considered when weighting the measured spectra.   

The measured spectra were corrected for the inefficiencies of the detection system, see Table~\ref{table:efficiencies}. The efficiencies of the scintillation detector and the Cherenkov detector were evaluated in a past experiment~\cite{Mihovilovic:2019jiz} and were determined to be $99.0\,\mathrm{\%}$ and  $99.85\,\mathrm{\%}$, respectively. The efficiency of the track reconstruction in the VDCs was determined to be $(99.98\pm0.05)\,\mathrm{\%}$.
All three corrections were considered as multiplicative correction factors. 

\begin{table}[h]
\caption{Detection efficiencies of the setup. The correction factors for the cross sections are given by their inverses. \vspace*{2mm}}
\label{table:efficiencies}
\begin{tabular}{@{}lll@{}}
\toprule
Contribution & Efficiency factor & Uncertainty \rule[-2mm]{0pt}{0pt} \\
\midrule
Scintillator efficiency & $0.990$ & $0.003$ \\
Cherenkov efficiency & $0.999$ & $0.001$ \\
VDC efficiency & $0.999$ &  $0.001$ \\
Particle identification & $0.983$  & $0.017$ \\
\botrule
\end{tabular}
\end{table}

\begin{figure}[!ht]
\includegraphics[width=1.0\linewidth]{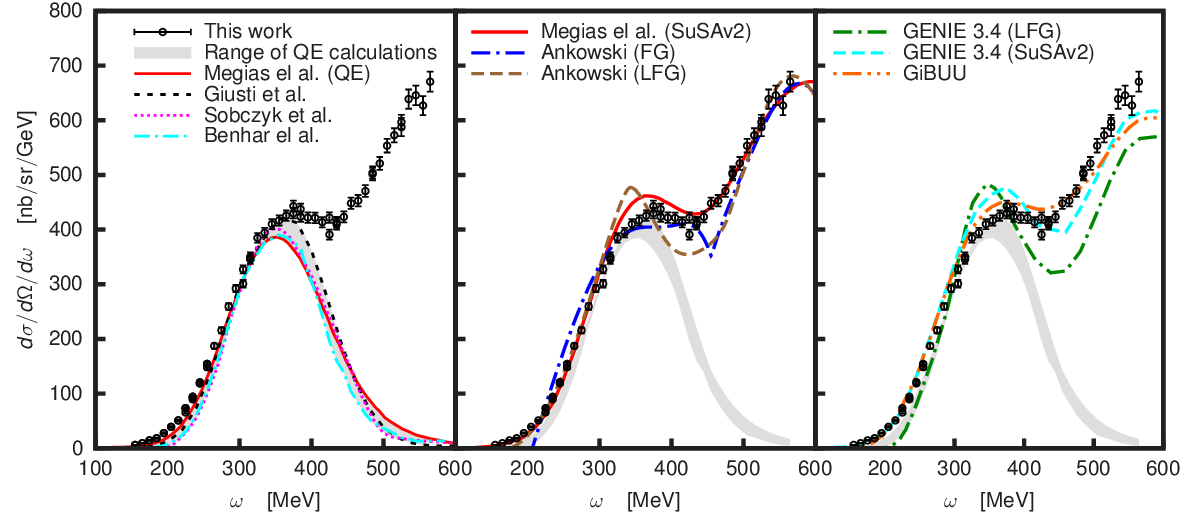}
\caption{
{\bf Left:} Measured cross section compared to the QE calculations of Megias {\it et al.} ~\cite{Megias:2016lke}, Giusti {\it et al.} ~\cite{Meucci:2003:PhysRevC.67.054601,Meucci:2009:PhysRevC.80.024605}, Sobczyk {\it et al.} ~\cite{Nieves:2017lij} and Benhar{\it et al.}~\cite{Benhar:1994hw}.
The gray band shows the envelope of the quasi-elastic cross-section calculations and represents a measure of the differences between different models.
{\bf Center:} Measured cross section compared to the full theoretical calculations of Megias {\it et al.} ~\cite{Megias:2016lke} based on the SuSAv2-MEC model and predictions of Ankowski~\cite{Ankowski:Private:2021} based on the global Fermi gas model (FG) and local Fermi gas model (LFG). {\bf Right:} Comparison of the new data with the results of the Monte-Carlo generators GiBUU~\cite{Buss:2011mx} and GENIE (version 3.4)~\cite{Andreopoulos:2009rq} employing LFG and SuSAv2-MEC~\cite{Megias_PhysRevD.101.033003} nuclear cross-section models.   
}
\label{fig_CS} 
\end{figure}

Several cuts were applied to both data and simulation. First, a cut on the Cherenkov signal was applied to identify electrons and minimize the background arising from cosmic particles and by negatively charged pions from the ${}^{12}\mathrm{C}(e,\pi^{-})$ reaction. This cut was followed by cuts on the nominal momentum and angular acceptance of the spectrometer in order to remove the artefacts at its edges, caused by the inefficient parts of the detectors, fringe fields in the spectrometers, and secondary particles rescattered from parts of the collimator. The extracted cross sections are  presented in Fig.~\ref{fig_CS}.
The systematic uncertainties of the extracted cross sections are a combination of various contributions. The uncertainties related to the detector efficiencies are collected in Table~\ref{table:efficiencies}. The uncertainty of the luminosity is given by the uncertainty of the absolute beam current calibration, which amounts to $3.3\,\mathrm{nA}$ at $855\,\mathrm{MeV}$ and the fluctuations of the beam current were related to the instabilities of accelerator operation. The latter were smaller than $5\,\mathrm{nA}$, resulting in the total systematic uncertainty smaller than $0.16\,\mathrm{\%}$. The dominant contribution to the systematic uncertainty is related to the misidentification of particles in the Cherenkov detector and cuts applied to distinguish electrons from pions and muons. This uncertainty was estimated to be $1.7\,\mathrm{\%}$. The last relevant contribution to the systematic uncertainty can be evaluated by the formalism of Mo and Tsai~\cite{MoTsai}, employed to describe radiative and multiple scattering corrections to the cross section.  These corrections add $0.2\,\%$ to the total uncertainty of the measured cross sections. Finally, the uncertainty of the position of the extracted cross sections on the energy scale is related to the ambiguities in the absolute energy calibration of the accelerator and spectrometer and amounts to $2.7\,\mathrm{MeV}$, which is less than $1/3$ of the employed energy bin size.  

\section{Comparison to models and event generators}

The extracted cross section is first compared to QE calculations of Giusti {\it et al.} ~\cite{Meucci:2003:PhysRevC.67.054601,Meucci:2009:PhysRevC.80.024605}, Sobczyk {\it  et al.} ~\cite{Nieves:2017lij}, Megias~{\it et~al.}~\cite{Megias:2016lke} and Benhar {\it et al.}~\cite{Benhar:1994hw}. Fig.~\ref{fig_CS}~(Left) shows that the calculations agree with each other at the level of $4\,\%$ at the top of the QE peak. The comparison of the experimental results to the comprehensive calculations of Megias~{\it et~al.}~\cite{Megias:2016lke}, which are based on the SuSAv2-MEC model, are shown in Fig.~\ref{fig_CS} (Center). The model exhibits very good overall agreement with the data, on average at the level of $7\,\%$. Surprisingly the largest inconsistency between the data and the calculations appears at the top of the QE peak, where the discrepancy is $9\,\%$. Since the QE calculations show a consistent picture there, the observed discrepancy between the data and the SuSAv2-MEC model is most likely related to the incomplete or inconsistent description of the processes in the ``dip" region, which are the only remaining relevant contributions to the cross section at $\omega \leq 400\,\mathrm{MeV}$. 

An agreement at a similar level has been achieved by Ankowski~\cite{Ankowski:Private:2021} who calculated cross sections by using both the global Fermi gas model (FG) and the local Fermi gas model (LFG) in combination with the Bosted-Christy~\cite{Bosted-Christy,Christy:2007ve} approach for describing pion production processes in the ``dip'' and in the $\Delta$-resonance region. The relative deviation of the FG calculation from the data is on average $10\,\%$, while the prediction of the LFG model agrees with the data at the level of $8\,\%$. Both calculations exhibit a visible inconsistency  at the top of the QE peak and in the ``dip" region, where the calculated cross-section do not follow the correct trend of the data.

Finally, the extracted cross-sections were compared also to the results of the Monte-Carlo generators GiBUU~\cite{Buss:2011mx} and GENIE~\cite{Andreopoulos:2009rq,Ankowski:2020:PhysRevD.102.053001}. Figure~\ref{fig_CS} (Right) shows that at the selected kinematic setting the generators describe the data reasonably well. GiBUU agrees with the data at the level of $9\,\%$. The accuracy of the GENIE generator depends on the model used for calculating nuclear cross sections. When the local Fermi gas model is used, the calculated cross section agrees on average with the data at the level of $22\,\%$. Similarly to the results of Ankowski's LFG model, the simulated cross section overestimates the QE cross section. Additionally, the GENIE simulation lacks strength in the $\Delta$-region, where the calculated cross section is $17\,\mathrm{\%}$ smaller than the data. The simulated results improve when GENIE uses the SuSAv2 model for describing QE scattering and processes in the ``dip" region. In this case we observe much better agreement between the data and simulation at the QE peak. Note that GENIE still uses its default model for describing the cross section in the $\Delta$-resonance region, which is the source of difference with respect to the Megias {\it et al.} result shown in the center panel.

\begin{figure}[!ht]
\includegraphics[width=0.8\linewidth]{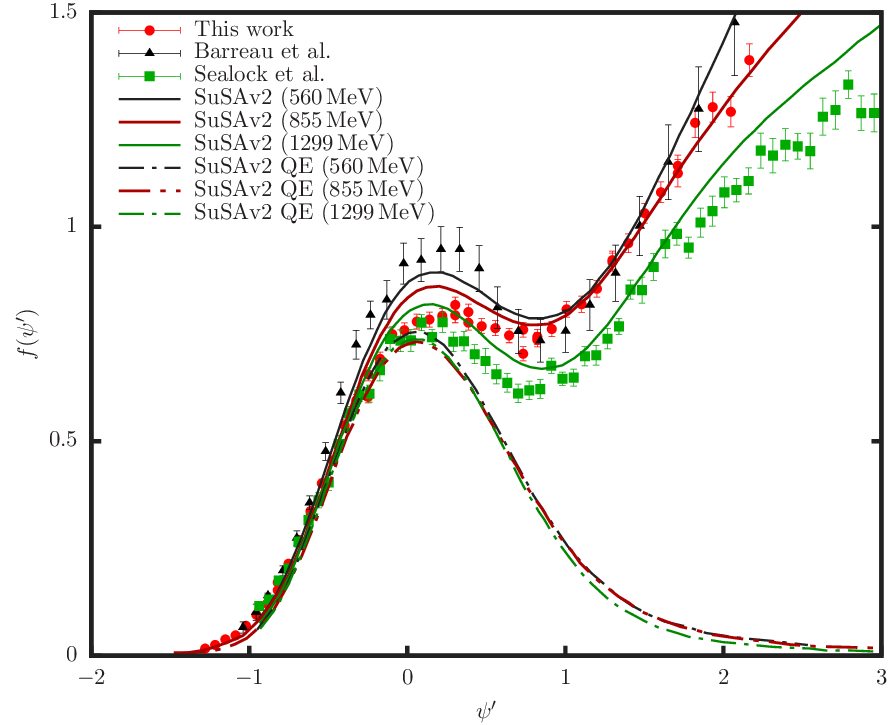}
\caption{The scaling function $f(\psi^\prime)$ at $|\vec{q}|\approx 0.8\,\mathrm{GeV}/c$. The experimental values of this work,  Barreau {\it et al.} ~\cite{BARREAU1983515}, and Sealock {\it et al.} ~\cite{Sealock1989} are shown together with the results of the SuSAv2-MEC model at corresponding beam energies of $560\,\mathrm{MeV}$ (Barreau {\it et al.}), $855\,\mathrm{MeV}$ (this work) and $1299\,\mathrm{MeV}$ (Sealock {\it et al.}) and presented with the full lines.  The dash-dotted lines are used to present contributions of the QE processes to the calculated scaling functions~\cite{Megias:2016lke}. 
\label{fig_Scaling} }
\end{figure}

For a more insightful analysis of the measured cross section and comparison with the existing results obtained under different kinematic conditions, the scaling formalism~\cite{Donnelly:1999sw} can be employed. The formalism was first developed within the framework of the relativistic Fermi gas model where the characteristic momentum is the Fermi momentum $k_F$, which can be expressed as a dimensionless scale parameter $\xi_F = \sqrt{1+ k_F^2/m_N^2}-1$. Building on this formalism, two dimensionless scaling variables $\psi$ and $\psi'$ were proposed~\cite{Maieron:2001it}:
\begin{eqnarray}
\psi &\equiv& \frac{1}{\sqrt{\xi_F}} \frac{\lambda-\tau}
{ \sqrt{ (1+\lambda)\tau + \kappa \sqrt{ \tau(\tau+1) } }}\,, \nonumber\\
\psi^{\prime} &\equiv& \frac{1}{\sqrt{\xi_F}}
\frac{ \lambda^{\prime }-\tau^{\prime } }
{ \sqrt{ (1+\lambda^{\prime }) \tau^{\prime } 
      + \kappa \sqrt{ \tau^{\prime }(\tau^{\prime }+1) }}
}\,. \label{eq:psiprime}
\end{eqnarray}
The variable $\psi^\prime$ is corrected for an empirical energy shift $E_\mathrm{shift}$ corresponding to the average of the separation energies of the various shells contributing to the nuclear ground state~\cite{Donnelly:1999sw}. With the shift one achieves the center of the QE peak to be at $\psi^\prime = 0$. The necessary shift is achieved by substituting $\lambda$ and $\tau$ with $\lambda^\prime = \lambda - E_\mathrm{shift}/2m_N$ and $\tau' = \kappa^2 - {\lambda^\prime}^2$.

The idea of the scaling formalism is to factorize the elastic cross section on a single nucleon, obtaining in this way a universal scaling function which contains information about the nuclear structure. 
For that purpose, reduced longitudinal and transverse response functions are introduced as~\cite{Maieron:2001it}:
$$
f_L  =  k_F~\frac {R_L}{G_L(\kappa,\lambda)}\,,\quad
f_T  =  k_F~\frac {R_T}{G_T (\kappa,\lambda)}\,.
$$
The functions $G_L$ and $G_T$ are expressed as:
\begin{eqnarray}
G_L(\kappa,\lambda) &=& \frac{ (\kappa^2/\tau) 
[ {\tilde G}_E^2 + {\tilde W}_2 \Delta ] }
{2\kappa [1+\xi_F (1+\psi^2)/2]}\,,\nonumber\\
G_T (\kappa,\lambda) &=& \frac{ 2\tau {\tilde G}_M^2 + {\tilde W}_2 \Delta } {2\kappa [1+\xi_F (1+\psi^2)/2]}\,, \nonumber
\end{eqnarray}
where
\begin{eqnarray}
\Delta &=& \xi_F(1-\psi^2)\left[
\frac{ \sqrt{\tau(1+\tau} } {\kappa} +\frac{1}{3} \xi_F (1-\psi^2) \frac{\tau}{\kappa^2} \right]\,,\nonumber\\
\tilde{W}_1 &=& \tau \tilde{G}_M^2\,,\nonumber \\
\tilde{W}_2 &=& \frac{1}{1+\tau} \left[\tilde{G}_E^2+\tau \tilde{G}_M^2 \right]\,,\nonumber \\
\tilde{G}_E^2 &=& Z~{G_E^p}^2 + N~{G_E^n}^2\,,\nonumber \\
\tilde{G}_M^2 &=& Z~{G_M^p}^2 + N~{G_M^n}^2\,.\nonumber
\end{eqnarray}
Here $Z$ and $N$ represent the number of protons and neutrons in the nucleus, respectively, $G_E^{p,n}$ and $G_M^{p,n}$ are nucleon electric and magnetic form factors~\cite{Kelly:2004hm}. Using these functions a dimensionless scaling function for the total cross section can be written as:
\begin{eqnarray}
f  &=& k_F \frac {d^2 \sigma/d\Omega_ed\omega}
{\sigma_M \left[ v_L G_L (\kappa,\lambda) +v_T G_T (\kappa,\lambda) \right]}\,.\nonumber\\
&=& f_L~\sin^2\chi_{TL} + f_T~\cos^2\chi_{TL}\,, \label{eq:f}
\end{eqnarray}
where the angle $\chi_{TL}$ is defined as
\begin{eqnarray}
\tan^2\chi_{TL} = \frac{v_L~G_L}{v_T~G_T} \label{eq:chiLT}\,.
\end{eqnarray}
This angle characterizes the ratio between the longitudinal and transverse contributions to the cross section. At $\chi_{TL}\approx 0$ the inclusive cross section is dominated by the transverse response, while at $\chi_{TL}\approx 90^\circ$ the cross section is governed by the longitudinal response~\cite{Donnelly:1999sw}.  

Using the scaling variables in Eqs.~(\ref{eq:psiprime}), (\ref{eq:f}), and (\ref{eq:chiLT}), the measured cross sections could be compared to the previous measurements at $|\vec{q}|\approx 0.8\,\mathrm{GeV}/c$ of Barreau  {\it et al.} ~\cite{BARREAU1983515} and  Sealock {\it et al.} ~\cite{Sealock1989}.  The extracted values of the dimensionless scaling function $f(\psi^\prime)$ are shown in Fig.~\ref{fig_Scaling}.  
This figure shows the approximate scaling of the measured cross sections which starts to break for $\psi'>0$ when the transverse contributions of the $\Delta$-resonance begin to dominate the cross section~\cite{Donnelly:1999sw,Maieron:2001it}. At the QE peak the experimental values of this work and Sealock {\it et al.} collected at scattering angles $\theta<90^\circ$, which corresponds to  $\chi_{TL} =36^\circ$ and $46^\circ$, respectively, agree very well with each other. On the other hand, the scaling function reconstructed from data of Barreau  {\it et al.} at $\theta = 145^\circ$ is over $20\,\mathrm{\%}$ higher at the top of the QE peak. These data coincide with a much smaller value of $\chi_{TL} = 12^\circ$, and are thus dominated by the transverse response $R_T$ which is known to break scaling due to various nonelastic contributions ranging from final-state-interaction (FSI) effects to contributions from the meson-exchange currents (MEC)~\cite{Donnelly:1999sw}. 

Experimental values were compared also to the full calculations of Megias {\it et al.}. Interestingly, at the top of the QE peak the theory is consistent with the data of Barreau  {\it et al.} , but overshoots the experimental values of this work and that of Sealock {\it et al.}. The analysis of the QE cross sections has revealed that for these two data sets the QE part matches the strength of the measured cross section, indicating that the discrepancy  might be due to the overestimated MEC contributions. 

\section{Conclusions}
We presented the experimental cross section for the inclusive reaction ${}^{12}\mathrm{C}(e,e')$ at $Q^2=0.8\,\mathrm{GeV}^2/{c}^2$. The measurement was made at the kinematics that is relevant for the accelerator based neutrino experiments, but where the available data are scarce.  Since the new data set has not been considered in any of the theoretical models, we could use them to challenge the calculations and generators employed in the interpretation of the experiments with neutrinos. We have demonstrated that the event generators in combination with selected nuclear models are capable of describing data at the level of $10\,\mathrm{\%}$. For even higher precision of the generators in the future, the built-it nuclear models need to be further refined, especially the description of the transverse part of the interaction, which governs the inclusive cross section in the region of the ``dip" and $\Delta$-resonance.
To achieve this goal, further theoretical and experimental investigations of cross sections at $Q^2\approx 1\,\mathrm{GeV}^2/{c}^2$ are needed. Kinematics at lower $Q^2$ would also be useful to test MEC models. In particular, the prospect of having ab-initio calculations in the light and mid-mass sector with MEC~\cite{Lovato12C_el, Seutin,Saori_A3, Acharya:2023ird,Miyagi:2023zvv,King:2024zbv} will motivate further experimental activities in the future on various targets. 

\backmatter

\bmhead*{Acknowledgements}
The authors would like to thank the MAMI accelerator group for the excellent beam quality which made this experiment possible. We also thank U. Mosel for useful discussions. This work is supported by the Federal State of Rhineland-Palatinate, by the Deutsche Forschungsgemeinschaft (DFG)  through the Cluster of Excellence ``Precision Physics, Fundamental Interactions, and Structure of Matter" (PRISMA$^+$ EXC 2118/1) funded by the
DFG within the German Excellence Strategy (Project ID 390831469), by the DFG grant "Electron Scattering on Nuclei for Neutrino Physics" (Project ID 521414474), by the Slovenian Research Agency under Grants P1-0102 and J1-4383, by the European Union’s Horizon 2020 research and innovation programme under the Marie Skłodowska-Curie grant agreement No.~101026014, by Croatian Science Foundation under the project IP-2018-01-8570 and by the University of Tokyo ICRR’s InterUniversity Research Program FY2024 (Ref. 2024i-J-001), Japan; the Spanish Ministerio de Ciencia, Innovaci\'on y Universidades and ERDF (European Regional
Development Fund) under contract PID2020-114687GB100 and by the Junta de Andaluc\'ia grant No. FQM160.

\bibliography{cque}

\end{document}